# Enhanced effective mass in doped SrTiO₃ and related perovskites

Wilfried Wunderlich[a,b,c], Hiromichi Ohta[a,b], Kunihito Koumoto[a,b],

a) Japan Science and Technology Agency, CREST, Kawaguchi 332-0012, Japan
b) Nagoya University, Graduate School of Engineering, Furo-cho, Chikusa-ku, Nagoya 464-8603, Japan
c) Tokai University, Department of Material Science Eng., Kitakaname, Hiratsuka 259-1292, Japan
E-mail: wi-wunder@rocketmail.com, Tel: +81-90-7436-0253

The effective mass is one of the main factors determining the Seebeck coefficient and electrical conductivity of thermo-electrics. In this *ab-initio* LDA-GGA study the effective mass is estimated from the curvature of electronic bands by one-band-approximation and is in excellent agreement with experimental data of Nb– and La– doped SrTiO₃. It is clarified that the deformation of SrTiO₃ crystals has a significant influence on the bandgap, effective electronic DOS- mass and band- mass, but the electronic effect due to the $e_{2g}$- band flattening near the Γ-point due to Nb -doping up to 0.2 at% is the main factor for the effective mass increase. Doping of La shows a linear decrease of the effective mass; this can be explained by the different surroundings of A- and B-sites in perovskite. Substitution with other elements such as Ba on the *A*-site and V on the *B*-site in SrTiO₃ increases the effective mass as well.

## Introduction

Heavy doping with Nb or La turns SrTiO₃ into an n-type degenerate semiconductor and a rather large thermoelectric figure of merit of 0.34 at 1000K is achieved[1-4], in superlattice structures even higher values due to the confinement in the two-dimensional electron gas.[5-9] Further improvement requires the increase of the effective mass as the most important factor[10] for the Seebeck coefficient and conductivity. This paper compares the experimentally determined values for the effective electron mass with those obtained from ab-initio calculations. In a previous study[1] an effective electron mass of $m^*/m_0$=7.2 ($m_0$=free electron mass) for SrTiO₃₋δ was determined from thermoelectric data, which increase with the Nb -doping concentration up to 0.2% and decrease with the La- doping concentration, and is consistent with $m^*/m_0$=15.[11] For very high Nb-concentrations of 5% at low temperatures $m^*/m_0$=1.8 and 1.9 for reduced for SrTiO₃ was measured.[12] A value of $m^*/m_0$=4.2 independent of the La-concentration in SrTiO₃ was determined[13] by interpretation of experimental data, while far smaller values of $m^*/m_0$=1.17 and 1.51 were reported for pure and La-doped SrTiO₃[14]. $m^*/m_0$=0.802 was obtained by ab-initio calculation for SrTiO₃.[15] Also spectroscopic measurements obtained an effective mass of $m^*/m_0$=5.0[16] for SrTiO₃ and $m^*/m_0$=3.3[17] for SrVO₃ Assuming that effective mass values $m^*/m_0$ obtained by different experimental measurement methods, like Hall-mobility $\mu$, conductivity $\sigma$ and Seebeck $S$ measurements,[18] are consistent, the discrepancy can be explained by the Kane model[17] which states that there is an increase of the effective mass $m^*/m_0$ with charge carrier concentration $n$, or in other words the bare electron mass[11] from bandstructure increases due to polaron effects. The dependence of these macroscopic properties $\mu$, $\sigma$ *and* S on the microscopic parameters $m^*/m_0$, $n$ and relaxation time $\tau$ has been proven by a semi-empirical theory[20] solving the Fermi-intergral. This approach for calculating the thermoelectric properties from $m^*/m_0$ was used in this paper and allows common ab-initio software[21,22] to be used straight-forwardly. Indeed such bandstructure calculations of the effective masses of heavy and light electrons and their average $m^*/m_0$ have been successfully performed for III-V semiconductors with excellent agreement with experimental data.[13] Other approaches have been performed, like calculating the thermopower directly from orbital interactions by solving Heikes' formula in the Hubbard model shown for the example of SrCoO₃.[21]

Ab-initio simulations on SrTiO₃[3, 14, 25-31] usually show a smaller band-gap than the experimental 3.2eV, but the difference vanishes, when a suitable self-energy $U$ is used (LDA+U calculations); nevertheless the main features of the bandstructure remain unchanged. The density- of- states (DOS) shows a remarkable steep increase at the valence band maximum (VBM) and the conduction band minimum (CBM) due to highly populated O-p- and Ti-$e_{2g}$-states.[26-30] Oxygen vacancies introduce a donor level a few 100meV below the CBM.[28-32] In general, large effective masses are caused by electron correlation effects near a metal-insulator transition (MIT).[33-35] In the case of La –doping, the band- insulator Sr₁₋ᵧLaᵧTiO₃ with $y$=0 becomes metallic near $y$=0.1 and again an insulator of Mott-type near $y$=0.9[33]. La occupies the Sr -site, and Nb the Ti -site, and in both cases the electron concentration increases linearly with the doping level, when the oxygen concentration is kept constant.[13,36] Nb-doping increases the valence of the Ti-site from 4+ to 5+, so that oxygen reduction becomes favorable and both defects[37] can only treated as independent for small concentrations. It is well accepted, that in these n-type Ti- perovskites only the electron conductivity needs to be considered, not holes. Furthermore, compared to other perovskites for lightly doped SrTiO₃ no clear evidence has yet been found for magnetic ordering, octahedron tilting or Jahn-Teller effects, although at highly doped SrTiO₃ these might occur.[29, 35, 38]

The goal of this paper is to find a reliable ab-initio method to calculate the effective mass which fits to experimental data of conductivity and Seebeck- coefficient measurements for both, doped and oxygen deficit SrTiO₃. First, the dependence of thermoelectric properties on the effective mass is shown. Then, the calculation method is explained, focusing especially how the effective mass from the electronic band structure and that from the density of states are related. The results of the bandstructure features and effective mass calculation are presented for several non-stochiometric, deformed, doped variants of SrTiO₃ as well as related substitutional perovskites. Finally, the influences of the different phenomena are summarized. In subsequent papers, the effective mass for layered perovskites was investigated.[3,7]

## 2. Calculation method

According to the equations in ref. 18, the Seebeck coefficient can be calculated phenomenologically by solving the Fermi integral if the effective mass m*, the carrier concentration n, and the scattering factor r=0.5 are known. For calculation of the electrical conductivity $\sigma$ the relaxation time for electron-phonon-scattering 10 is also required, which was found to be independent of La- and Nb- doping1 for SrTiO3, $\tau$= 18 fs at T=300K. In previous calculations3 by using an empirical method20 the parameters m*, n were varied over wide ranges and the resulting negative Seebeck coefficient −S, and conductivity $\sigma$ showed excellent agreement to experimental data.1 The Seebeck coefficient |−S| increases steeply when the effective mass increases from $m^*/m_0$=1 to 5 and moderately above that value, while the conductivity decreases due to smaller mobility. The resulting power factor $S^2 x \sigma$ as a function of m* and n shows that the decrease in mobility by increasing m* is almost compensated by the gain in S2, while the charge carrier concentration indeed requires optimum adjustment between the insulating and metallic behavior. Such calculations can be used to estimate m* and n from S and $\tilde\sigma$ Room temperature data on m* of doped SrTiO3 are summarized in Table 1 and show good agreement among each other.

The ab-initio software program Vasp21 based on the density-functional theory (DFT) with generalized gradient approximation (GGA) was used in this study with a pseudo-potential cutoff-energy of $E_{cutoff}$= −380eV. The Wien2k22 code which uses the linear augmented plane-wave (LAPW) method leads to the same results, respectively. Both ab-initio codes include the relativistic approximation and the spin-orbit interaction and show excellent agreement with previously published data.[14,25-31] The crystallographic data of SrTiO3 (Pm-3m, $a_0$=0.3905nm) are used in general, but as the experimental lattice constants of the substrate-constraint films were found to be larger ($a$=0.3930 nm),2 calculation with varying lattice constants from 0.371 to 0.402nm ($a$=0.95 $a_0$ to 1.03 $a_0$) was also performed. For modeling the A - and B -site substitution as well as the strained SrTiO3 lattice, simple unit cells were used, while for modeling the doping or vacancies 2x2x2 extended supercells were used, in which the atoms around the defect were allowed to relax by the ab-initio molecular dynamic option within the Vasp code.21 The lattice constants for doped supercells were adjusted assuming a linear increase according to Vegards rule using for experimental values a=0.4107nm29 for SrNbO3 and a=0.3935nm17 for LaTiO3. Each of the eight Sr- atoms was step-by-step substituted randomly by La or the Ti-atoms by Nb. Further calculations were performed by leaving one of the 24 O-atoms vacant, and additionally calculations with both defects, Nb-doping and Oxygen vacancies were performed. The supercell limitation leads to a restriction of a minimal doping concentration for the calculation, namely SrTi$_{1-x}$Nb$_x$O$_{3+v}$, or Sr$_{1-y}$La$_y$TiO$_{3-v}$ of x,y >0.125 and v>0.042, while the doping concen-

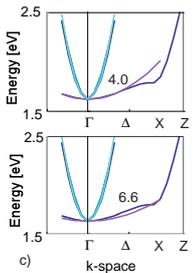

**FIG. 1.** Effective mass estimated from fit parables (bright lines, eq. (1)) to the calculated (dark lines) band structure of the conductions band of SrTiO3 near the Γ-point in Γ−Δ−direction

**Table 1.** Experimental thermoelectric properties of doped SrTiO3

| Doping element | Doping concentration x or y [at-%] | Carrier concentration n [$10^{20}$cm$^{-3}$] | Effective mass $m^*/m_0$ | Seebeck coefficient S [µV/K] | Conductivity $\sigma$ [S/cm] | Ref. |
|---|---|---|---|---|---|---|
| La | 0.005 | 0.5 | 6.6 | -420 | 54 | 1 |
| La | 0.07 | 6.8 | 6.0 | -150 | 1000 | 1 |
| Nb | 0.01 | 1.5 | 7.3 | -330 | 95 | 1 |
| Nb | 0.02 | 3.7 | 7.7 | -240 | 353 | 1 |
| La | 0.02 | 4.0 | 4.2 | -200 |  | 3 |
| La | 0.02 | 3.7 | 4.5*) | -260 | 433 | 15 |

*) Instead of the published value of 1.51 in ref. 15, a value of $m^*/m_0$=4.5 calculated by the equations in ref. 20 by relating n, and m* to S, and $\sigma$ leads to data consistency.

tration in experiments is one order of magnitude smaller (table 1).

The effective mass obtained from the bandstructure $m_{B,i}$ in the following called effective band mass, was estimated in the one-band-approximation from the curvature of each band i at each extreme points in k-space j (usual Γ point); or in other words, parts of the band mass $m_{B,ij}$ for each band i at each j-point, conduction band minimum (CBM) or valence band maximum (VBM), were fitted by parabolic curves as

$$\frac{1}{m_{B,i,j}} = \frac{2}{\hbar^2} \cdot \frac{d^2 E_{i,j}}{dk_{i,j}^2}$$

(1),

with Planck's constant $\hbar$. The averages over j, namely heavy (h) and light (l) parts from the same band i were calculated according to the following formula from ref. 41 yielding to the effective band mass

$$m_{B,i} = \left( m_{B,i,h}^{3/2} + m_{B,i,l}^{3/2} \right)^{2/3}$$

(2).

By comparison with experimental data and the above mentioned observation1,2 that in n-semiconductors only electrons are relevant for electric conductivity, it was found that the mass from electrons with the lowest energy in the conduction band determines the electrical transport properties and is equivalent to the effective mass m* in Drude's equation, which is measured in experiments, can be calculated from

$$m^* = m_e^* = m_{B,i}$$

(3)

with $i$= 1 (minimal conduction band). The main statement of this paper is that formulas (1-3) allow the estimation of effective mass m* measured in experiments by ab-initio bandstructure calculations from $m_{B,i}$. The values for m* derived by this method showed good agreement with literature values of the effective masses for both electrons and holes at all significant reciprocal space points of the III-V semiconductors GaN, GaP, GaAs, InAs and InAs, as well as ZnO and TiO2.

One important remark should be added. In literature the effective electron mass m* is sometimes set equivalent to the DOS mass $m_{DOS}$. In order to prove this, the DOS masses $m_{C,DOS}$, $m_{V,DOS}$ for either the conduction (C) and valence (V) band were estimated by fitting the envelope of the density of states with the following functions $D_C(E)$ or $D_1(E)$ for conduction- and valence- bands below $E_{CBM}$ (energy of conduction band minimum) or above $E_{VBM}$ (energy of valence band maximum):

$$D_C(E) = \frac{V}{2\pi^2} \cdot \left(\frac{m_{C,DOS}}{\hbar^2}\right)^{3/2} \cdot (E_{CBM} + E)^{1/2}$$

$$D_V(E) = \frac{V}{2\pi^2} \cdot \left(\frac{m_{V,DOS}}{\hbar^2}\right)^{3/2} \cdot (E_{VBM} - E)^{1/2}$$

(4).

By taking the arithmetic average of the effective band masses $m_{B,i}$ for several bands (n > 3) above the band-gap the effective DOS mass can be related to the effective band mass $m_{B,i}$

$$m_{DOS} = \frac{1}{n} \sum_{i=1}^{n} m_{B,i}$$

(5).

The $m_{DOS}$ obtained by averaging over effective band mass according to equation (5) and that derived from the density of states by equations (1,2) show excellent agreement to each other. However, it should be mentioned that $m_{DOS}$ is not related to the experimentally determined effective mass $m^*$, as can be seen for example, on the opposite dependences of $m^*$ and $m_{DOS}$ on the Nb-content for doped SrTiO$_3$ described in the last section (fig. 5). This means, that unfortunately the less costly calculation of the density- of- states cannot be used to estimate the effective mass $m^*$ relevant for electric conduction properties; only that from band structure calculations can be used.

The experimental data of $m^*/m_0$ for doped SrTiO$_3$ are summarized in Table 1 and vary from 4.2 to 7.2. The calculations for pure SrTiO$_3$ show a value of $m_{B,i}/m_0 = 4.4$ for heavy electrons around the $\Gamma$-point in the $\Gamma-\Delta$ (100) direction, while the masses in the $\Gamma-\Sigma$ (110) or $\Gamma-\Lambda$ (111) directions are smaller ($m_{B,i}/m_0 = 1.1$), yielding $m^*/m_0 = 4.8$ which is in good agreement with experimental values[1,3,15] (Table 1) When comparing calculations of a single unit cell and a 2x2x2 supercell of undoped SrTiO$_3$ the band- curvature, and hence the effective mass, is unchanged; the only difference is that the degeneracy is suppressed. While $m_{B,i}/m_0 = 4.4$ is obtained when fitting in the vicinity of the $\Gamma$-point (Fig. 1 (a)), a larger effective mass of $m_{B,i}/m_0 = 6.6$, resulting in $m^*/m_0 = 6.8$, is obtained (Fig. 1 (b)), when the fit is extended towards X including non-parabolic deviations, namely a small hump in the difference between the band and the parabolic fit, which appears in highly-accurate Vasp calculations only. In the following the parabolic values without the hump were used, but note that this non-parabolic behavior is one of factors explaining the large effective mass of 7.2 in the experiment. Further factors, such as lattice expansion, oxygen vacancies and doping as described later, lead to a further increase of $m^*$ and when considering all contributions, $m^*/m_0$ matches indeed the experimental value of 7.2[1] (Table1). On the other hand, the higher purity of SrTiO$_3$ in ref. 13 might be the reason for their small effective mass $m^*/m_0 = 4.2$, or the corrected value of 4.5 in ref. 15.

## 3. Results

### Effective masses of distorted SrTiO$_3$ lattices

The calculated electronic band structure of SrTiO$_3$ shows excellent agreement with literature data[4,15-17] with its slightly indirect band gap due to the higher valence band level at the $\Sigma$ and M points compared to the $\Gamma$ point. As most other researchers did[14-16], we will in the following renounce the plot of the M-R section[25]. When the SrTiO$_3$ lattice is isotropically expanded ($a,b,c=1.1a_0$) (Fig.. 2(a)) the lowest conduction band near the $\Gamma$ point in the $\Delta$-(100)- direction shows a smaller curvature, and hence the highest effective mass $m_B/m_0 = 5.0$, compared to 1.2 in $\Sigma$- direction and 1.3 in the $\Lambda$-direction, yielding an average value according to eq. (2) of $m^*/m_0 = 5.8$. While the band structure of the isotropic expanded lattice does not change its morphology (Fig. 2(a)), an orthorhombic distortion ($a=0.375$nm, $b=0.4066$nm) shows a suppression of the degeneracy, in other words band splitting occurs (Fig.. 2(b)). Although there is still a conduction band (i=3 counting from $E_{CBM}$) with a large effective mass 3.6, the lowest band in energy (i=1, see Table 2) now has a strong curvature, which leads to $m^*/m_0 = 0.3$. The three-dimensional plot in Fig. 3 summarizes the dependence on the lattice constants a, b for the unit cell volume (Fig. 3(a)), the total energy (b), the bandgap (c) and the effective mass $m_{B,i}$(d). Along the line A-A (Fig.. 3(a)) the lattice is strained isotropically ($a=b$), as shown in the upper left inset, while along B-B the volume is constant, as shown in the right inset. The lattice constant c was determined by $c=a*b/a_0^3$. The distortion of epitaxially grown thin films with $a$ and $b$ adjusted to the substrate, leaving $c=a_0$ unchanged, would be an intermediate case (middle inset). The band-gap decreases steeply from 3.2eV of undeformed SrTiO$_3$ to 2.5eV when the lattice is expanded ($a,b,c=1.03a_0$), while the effective mass increases to 8.2 (see also Table 2). The band-gap decrease can also be recognized in the density- of- states plot as a function of lattice constants in Fig. 4(a). This plot also clearly shows the increase in $m_{DOS}$ due to steeper envelopes (see also Table 2). Fig. 3 and 4(b) are symmetric along the axis A-A. When the lattice is distorted in orthorhombic shape, the bandgap is almost unchanged as derived from the DOS (Fig. 4(b)). Also the band effective mass is unchanged ($m^*/m_0$=4.5) as seen in Fig. 4(d). The effective mass $m_{DOS}$ decreases, because band splitting smoothens the DOS distribution. In summary, the effective mass can be increased by a factor of 1.7 when the lattice is expanded by a factor of 1.1, but the expansion needs to be isotropic. These lattice constant expansion studies require less time-consuming calculations than the following studies of stochiometry deviations or doping, but are much more difficult in experiments.

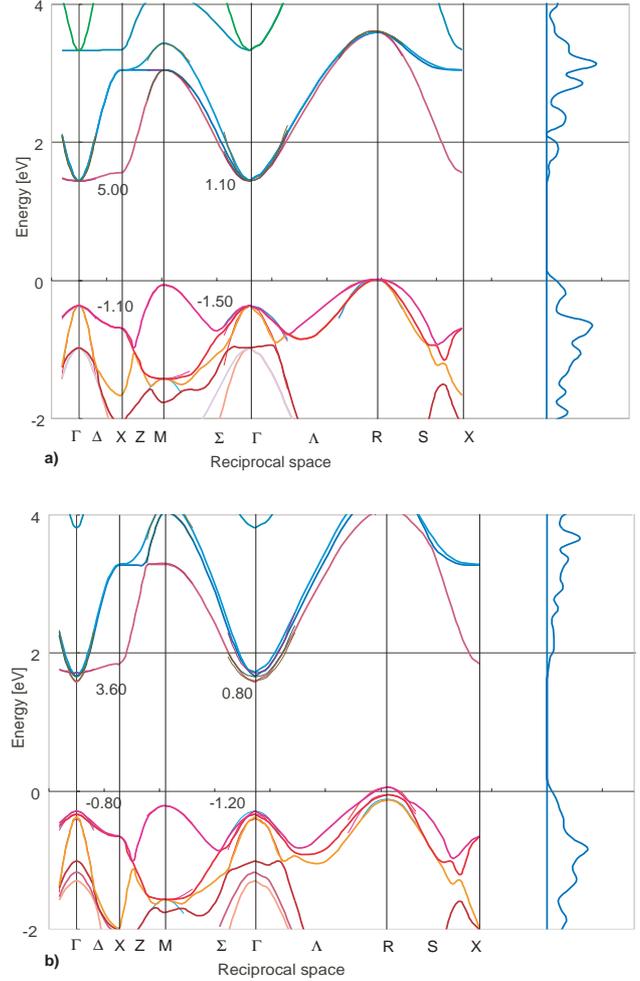

**FIG. 2.** Electronic band structure and DOS of deformed SrTiO$_3$ unit cells (a) $a=b=c=0.402$nm, (b) $a=0.375$nm, $b=0.4066$nm with the effective mass for electrons and holes.

**Table 2** Calculated effective masses $m_{DOS}$, $m_{B,l,j}$, and $m^*$ of doped or strained SrTiO₃. Values are in units of the free electron rest mass $m_0$. The values for $\Gamma-\Sigma$, or $\Gamma-\Lambda$ are summarized in one column, because they are almost identical.

| Composition | Valence band | | | | Conduction band | | | |
|---|---|---|---|---|---|---|---|---|
| | $m_{DOS}$ | $m_{B,l,j}$ | | $m^*$ | $m_{DOS}$ | $m_{B,l,}$ | | $m^*$ |
| | | $\Gamma-\Delta$ | $\Gamma-\Sigma$, or $\Gamma-\Lambda$ | | | $\Gamma-\Delta$ | $\Gamma-\Sigma$, or $\Gamma-\Lambda$ | |
| SrTiO₃ near Γ (Fig. 1(a)) | 4.59 | 3.5 | 2.8 | 9.8 | 3.53 | 4.4 | 1.1 | 4.8 |
| SrTiO₃ Γ-extended (Fig. 1(b)) | 4.59 | 3.5 | 2.8 | 9.8 | 3.53 | 6.6 | 1.1 | 7.5 |
| SrTiO₃, $a$, $b$, $c$ =1.1$a_0$ | 5.17 | 3.5 | 2.7 | 9.4 | 5.17 | 6.3 | 1.3 | 8.2 |
| SrTiO₃, $a$, $b$, $c$ =0.9$a_0$ | 4.81 | 3.5 | 2.7 | 9.4 | 2.81 | 2.9 | 1.0 | 2.9 |
| SrTiO₃, $a$=0.9$a_0$, $b$=1.1$a_0$, i=3 | 4.96 | 3.5 | 2.8 | 9.8 | 2.83 | 3.6 | 0.9 | 3.4 |
| SrTiO₃, $a$=0.9$a_0$, $b$=1.1$a_0$, i=1 | 4.96 | 3.5 | 2.8 | 9.8 | 2.83 | 0.3 | 0.9 | 0.3 |
| SrTiO₂.₉₆ | 3.62 | 10 | 2.7 | 27 | 2.9 | 4.4 | 1.8 | 7.7 |
| SrTiO₂.₉₂ | 2.60 | 8.0 | 2.05 | 16 | 2.23 | 2.0 | 1.6 | 7.2 |
| SrTi₀.₈₈Nb₀.₁₂O₃ | 3.25 | 10 | 2.7 | 27 | 2.17 | 6.0 | 1.1 | 6.3 |
| SrTi₀.₇₅Nb₀.₂₅O₃ | 2.96 | 4.0 | 3.0 | 12 | 2.07 | 0.2 | 1.0 | 0.2 |
| SrTi₀.₈₈Nb₀.₁₂O₂.₉₆ | 3.99 | 0.6 | 1.6 | 0.9 | 3.99 | 4.0 | 1.2 | 4.8 |
| SrTi₀.₈₈Nb₀.₁₂O₂.₉₂ | 2.96 | 0.4 | 1.5 | 0.6 | 3.5 | 8.0 | 1.3 | 7.8 |
| Sr₀.₈₈La₀.₁₂TiO₃, i=1, a= $a_0$ | 4.76 | 10 | 2.7 | 27 | 3.74 | 3.6 | 1.1 | 3.7 |
| Sr₀.₈₈La₀.₁₂TiO₃, i=1, a=0.99 $a_0$ | 4.85 | 4.0 | 1.4 | 4.9 | 3.81 | 3.6 | 1.2 | 4.3 |
| Sr₀.₇₅La₀.₂₅TiO₃, i=2, a=0.98 $a_0$ | 4.42 | 6.0 | 2.6 | 13 | 3.98 | 3.2 | 1.3 | 4.2 |
| Sr₀.₇₅La₀.₂₅TiO₃, i=1, a=0.98 $a_0$ | 4.42 | 6.0 | 2.6 | 13 | 3.98 | 0.2 | 0.7 | 0.2 |

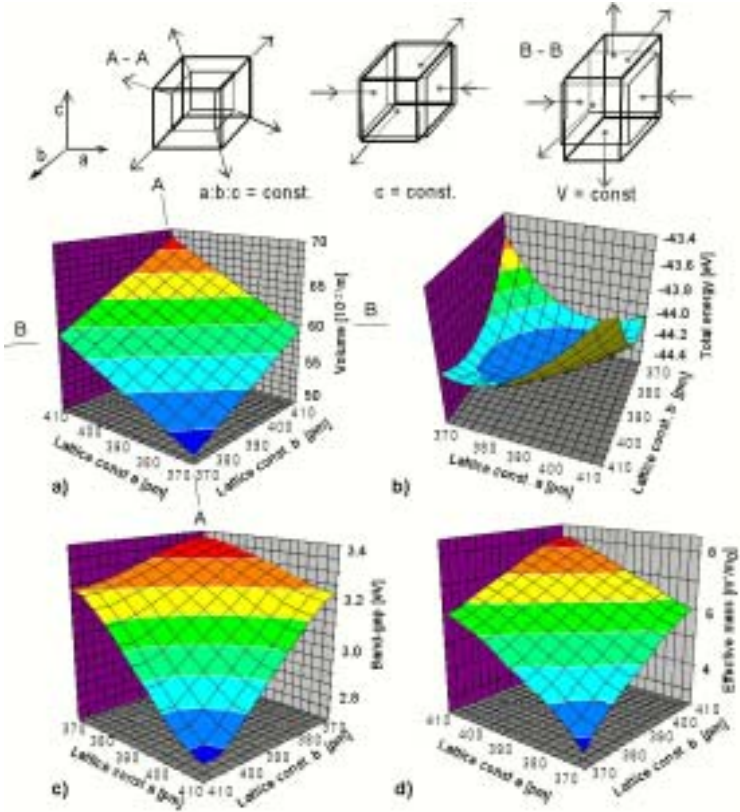

**FIG. 3.** (a) Volume, (b) total energy, (c) bandgap, (d) effective mass $m^*/m_0$ of deformed SrTiO₃. The inlet shows the deformation states, along the line A-A marked in (a) isotropic expansion, along B-B orthorhombic distortion with constant volume.

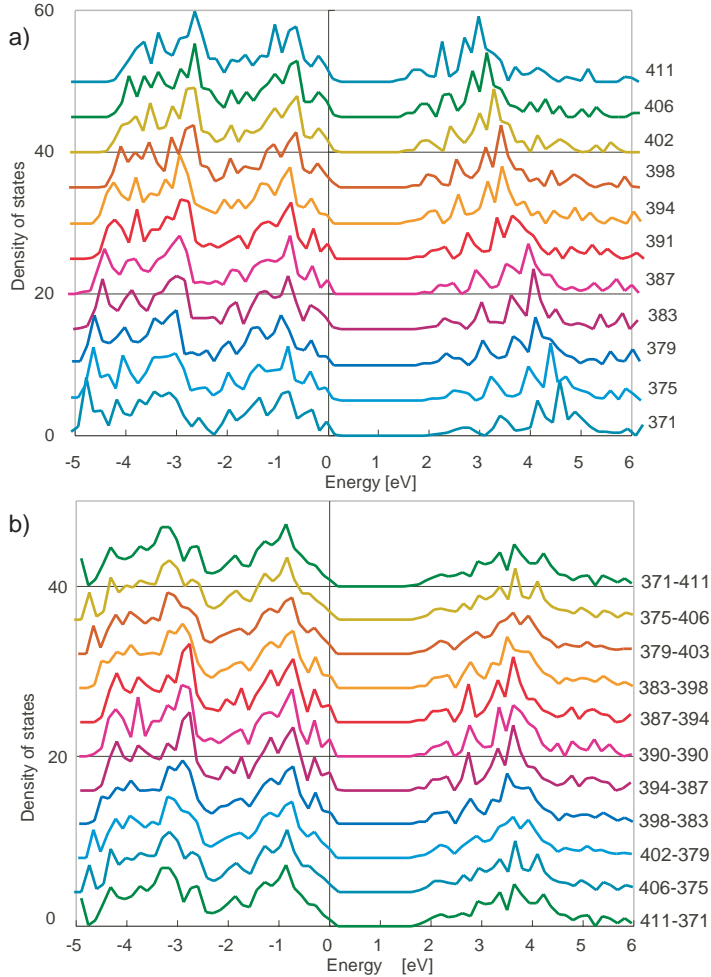

**FIG. 4.** Total density of states of deformed SrTiO₃, (a) isotropic expansion (line A-A in fig. 3a), (b) orthorhombic distortion with constant volume (line B-B in fig. 3(a)). The numbers refer to the lattice constants in pm.

### Lattice Vacancies

While pure SrTiO₃ is an insulator, doping with electron donating ions like $La^{3+}$ for $Sr^{2+}$ or $Nb^{5+}$ for $Ti^{4+}$ turns SrTiO₃ into a n-type semiconductor. To maintain the charge balance the material needs to compensate the surplus charge by introducing oxygen vacancies, in other words, in order to obtain semi-conducting doped SrTiO₃, it requires sintering in oxygen reduced atmosphere turning the white color of insulators into black. In this section we describe the oxygen reduction and in the next section the influence of doping. In the case of SrTiO₂.₈₇₅, which is the smallest concentration to be calculated by removing one oxygen atom in a 2x2x2 supercell, the electronic bandstructure shows an additional donor band about 300meV below the usual conduction band as shown in Fig.10 e and in Fig. 3 of ref. 18. Concerning the curvature directly at Γ in the Γ-Δ direction, the effective mass is almost the same as for pure SrTiO₃ ($m_B/m_0$=4.4), but compared to other bands it has less curvature in the Γ-X or Γ-R direction with a higher average effective mass ($m*/m_0$=1.8) than pure SrTiO₃ (Table 2). Including the considerations of Fig.. 1(b) about the non-parabolic behavior are taken into account, the mass is remarkably larger ($m*/m_0$>10). However, when the oxygen content is lowered further (SrTiO₂.₇₅, see Table 2), the additional band becomes strongly deformed and the mass decreases ($m*/m_0$<2), as well as $m_{DOS}$. This result confirms that electronic effects strongly depend on the oxygen concentration range. Vacancies on the A- and B-sites were also studied. Low concentrations of vacancies on the Sr-site (Sr₀.₈₇₅TiO₃) increases the effective mass slightly ($m*/m_0$= 4.9), and at higher concentrations Sr₀.₇₅TiO₃ even more $m*/m_0$= 9. Vacancies on the Ti -site increase the effective mass slightly. In summary, any vacancies in small concentrations increase the effective mass.

### Doping with La on A- site or Nb on B- site

The doping concentrations necessary for turning SrTiO₃ into a n-type semiconductor in experiments are smaller than those achievable by present calculations (x>0.12), but the following results shows the trend. Substitution of Ti- by Nb- atoms in a 2x2x2 supercell as in SrTi₁₋ₓNbₓO₃ increases the effective band mass $m*$ steeply in both, the valence band ($m*/m_0$> 20, Fig. 5(d)) and conduction band ($m*/m_0$> 8, Fig.. 5(c)). This increase in effective mass was confirmed also in 5x1x1 and 3x2x1 supercells for x < 0.25. Above x = 0.25, the effective mass suddenly drops to a low value of $m*/m_0$<3. The DOS-mass, however, shows a smooth dependence on x (Fig.. 5 (a,b)), while the bandgap decreases from 3.2 to about 2.5eV until x = 0.5 and increase again (Fig.. 6(a)). On the other hand, the doping on the B- site, Sr₁₋ᵧLaᵧTiO₃, shows a continuously increasing band gap (Fig.. 6(b)) with y and decreasing effective mass (Fig.. 7), which will be explained later. The unusual increase of the effective band mass becomes clearer, when the details of the bandstructure of SrTi₁₋ₓNbₓO₃ with x=0.12 and 0.25, before and after the discontinuity in $m*$, are compared (Fig.. 8). The double degenerated bands with large and small masses $m_B/m_0$= 4 and 0.2 in SrTiO₃ split at the Γ point in energy with a difference of 0.15eV. Between X-Z and X-R they are still degenerated, but becomes also suppressed at x=0.12 leading to the increase in $m*$. At x=0.25 (Fig.. 5(b)) the energy degeneracy at Γ between the bands with large and small masses vanishes and at x=0.38 also the degeneracy between X-Z and X-R is lost, although it reappears above x ≥ 0.75. In the valence band a similar suppression of the degeneracy near Γ occurs above 0.25, but above x ≥ 0.75 the topology of bands is different than below x < 0.25, and causes the unsymmetrical behavior in Fig.. 5(c) and (d). In summary, we can state that the large increase in effective mass is caused by degeneration in energy between a heavy and a light conduction band, both with lowest energy. This degeneration cannot be held at higher doping concentrations and the light band becomes lower in energy than the heavy one leading to a smaller bandgap and small $m_{DOS}$ between x = 0.25 and 0.75. In this concentration range the $m*$ relevant for macroscopic properties may become even smaller (0.2, as shown in Table 2) than shown as $m_{C,h}$ in Fig.. 5 (c), because the band with the light mass $m_{C,l}$ is now lower in energy (band with i=1) between x=0.25 and 0.75. This concentration range is distinguished from the usual case of a heavy conduction band at lowest energy, with the open symbols in Fig.. 5(c).

Such behavior in the bandstructure occurs similarly for Sr₁₋ᵧLaᵧTiO₃, the La-doping on the A-site, namely the splitting of a degenerated SrTiO₃-band into a light band and a heavy band with increasing y. At the same concentration as in the Nb-case, y=0.25, the light band is shifted about 40meV below the heavy band, marked with open symbols in Fig.. 6, which shows the average effective mass m*. At the same time the effective band masses decreases to $m_B/m_0$ =3.6, which leads to a smaller average effective mass than SrTiO₃, $m*$=3.7 or 4.3 (Table 2, Fig. 6) for $a = a_0$ or $a = 0.99a_0$, respectively. This decrease in effective mass is smooth, different from the Nb -case and was also observed in the experiments[2]. The density-of-states envelope has almost the same shape for all concentrations (Fig. 6(b)), yielding a nearly constant $m_{DOS}$ = 3.85. However, the bandgap increases with y from 3.2 to 4eV. In summary, La -doping decreases the effective mass almost linearly, while Nb -doping increases the mass steeply until the separation of the heavy and light bands causes a drastic decrease.

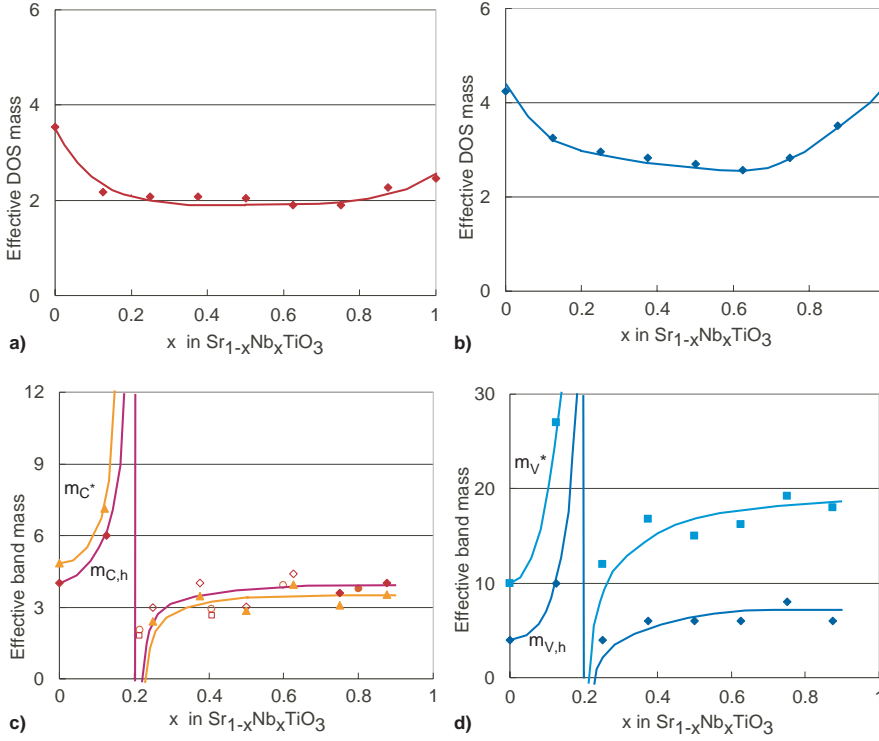

**FIG. 5.** Effective mass as a function of the Nb- composition x in SrTi₁₋ₓNbₓO₃. (a,b) show the effective DOS-mass for the (a) conduction band, (b) valence band, (c,d) effective band mass $m*$ (bright line) and $m_{B,h}$ (dark line) for (c) conduction band, and (d) valence band. The lines are eye guides. Filled and open symbols refer to heavy and light bands at lowest energy; rhombic, square and round symbols refer to 2x2x2, 5x1x1, and 3x2x1 supercell-calculations, respectively.

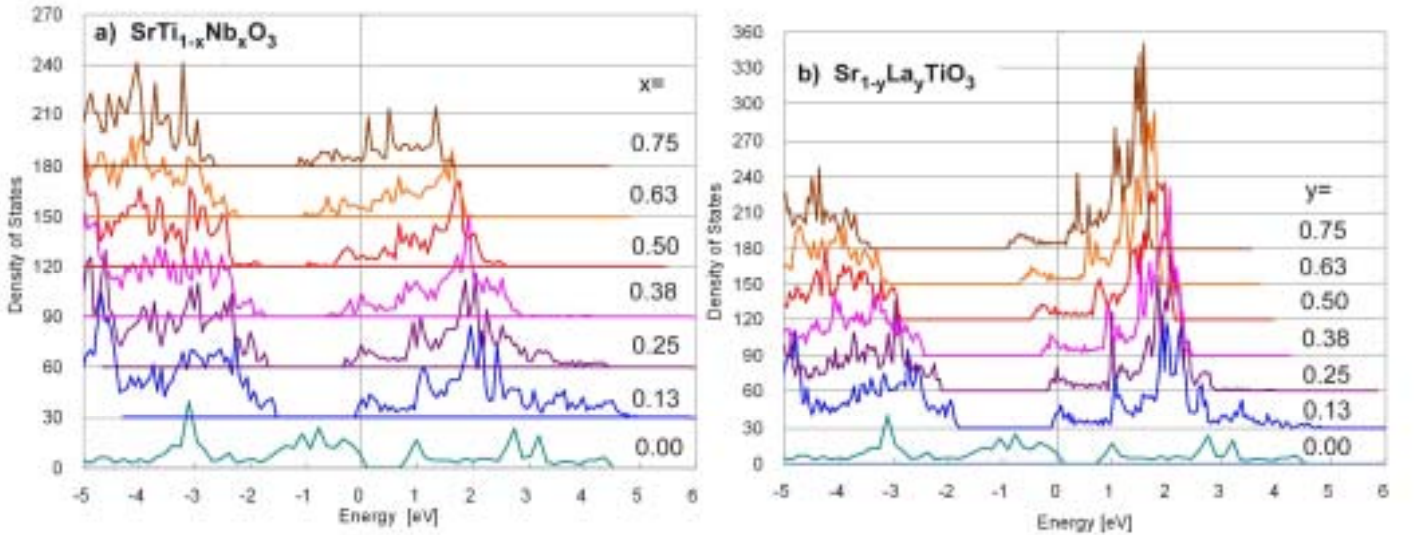

**FIG.6.** Density of states for (a) Nb-doped, (b) La-doped SrTiO₃

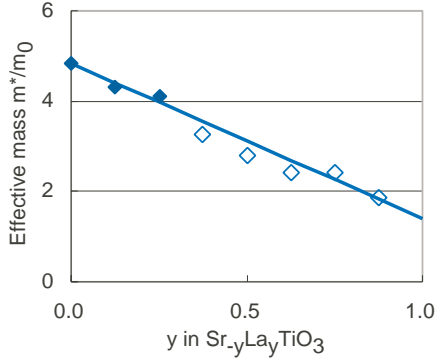

**FIG. 7.** Effective band mass $m^*$ for La-doped $SrTiO_3$ . (Dark symbols i=1, bright i=2).

### Substitution of other elements on A- or B-site

When other elements are substituted on the *A*- or *B* -site in $SrTiO_3$-related perovskites, the lattice constants change slightly but the morphology of the band structure remains almost the same. The change in band curvature leads to values of the effective masses shown in Table 3. The effective masses $m_{B,I}(\Gamma-\Delta)$ and $m^*$ increase in the order *A*= Ca, Sr, Ba according to the increase in atomic mass for elements in the same column of the periodic table. On the *B* -site, however, it decreases in the order *B* = Ti, Zr, Hf. For elements with different valence than $Ti^{4+}$ no simple rule is obtained, except that low valence (+2 and +3) leads to smaller effective masses as in the case of Cu, Co, Fe and Mn. Whether the opposite conclusion, high valence on the *B* -site lead to large masses, is true, cannot clearly be answered yet, but in the case of heavy $Ta^{+5}$, the charge exchange is already so strong that in order to stabilize the perovskite structure, the *A*- site would need to be occupied by valence (+1)-elements (Table 3). These compounds seem to have fairly high effective masses, while for the half- *A* -site occupied $Ca_{0.5}TiO_3$ the effective mass becomes rather light.

In order to compare the effective mass dependence on Nb- and La- doping, the lattice parameter was chosen. It increases linearly with Nb- concentration and decreases with La- concentration. The experimental data of ref. 1 and 2 are shown in Fig. 9 for the single crystal (squares) and thin films (triangle). In order to compare these values with the calculations, the lattice constant of pure $SrTiO_3$ (0.3905nm) taken in the calculations was adjusted to the experimental values marked with $a_{exp}$ as well as the effective band mass to the experimental value $m^*/m_0$=7.2 marked with the hatched lines and the round dot in Fig. 9. This calibration avoids the difficulties caused by impure samples and non-parabolic bands. The increase of the effective mass with increasing lattice constant from Fig. 4(d) is shown as a dotted line. The bright line in Fig. 9 refers to the steep increase of $m^*/m_0$ with the Nb-concentration according to Fig. 5(c) as well as the linear decrease with La-concentration from Fig. 6. Except the $SrTi_{0.6}Nb_{0.4}O_3$ thin-film specimen, all other data point fit the graph perfectly. The calculated increase of the effective mass, when the lattice is expanded, is shown in Fig. 9 as a dotted line. Hence, the geometric contribution can be separated from the electronic contribution, which increases steeply at *x*=0.2 for Nb- doping.

### Defect Interaction: Doping and Oxygen vacancies

As described above, in order to keep the charge balance the n-doping in experiments is associated always with the formation of oxygen vacancies, but the vacancy concentration in experiments is much smaller than in simulation, otherwise the supercells for modeling are getting very large. Fig. 10 summarizes the main features of different simulated cases, a) no doping b) increased effective mass for x=0.12, c) 0.24, d) 0.37, which has been summarized in fig. 5d: leads to the steep increase of the effective mass until x=0.25, but at x>0.25 it leads to the break-through of a light band with very small mass $m^*/m_0$=0.2. When vacancies and lattice relaxation around them are introduced $SrTi_{1-x}Nb_xO_{3-v}$ (fig. 10 e), g) v=0.12, or f), h) v=0.24), a new band 300meV below the valence band minimum occurs, but the band curvature becomes complicated (fig 10 e-g). In order to explain the small doping concentration and large vacancy dilution in the experiment (x=0.02, and v<0.2) l we suggest here that each defect can be treated as if separated. Hence, the bandstructure turns into quasi-p-n junctions as sketched in fig. 11b. The 300meV band turn into donor levels below the Fermi-energy and traveling electrons maintain their large effective mass when passing through a defect. This can explain why a large effective mass is measured in the experiment $SrTi_{1-x}Nb_xO_{3-x}$, with x=0.02, while in simulation a combination of Nb-doping and oxygen vacancies leads to small effective masses (table 2). The explanation is that the one-band approximation treated here is valid for nearly free electrons, as in pure $SrTiO3$, where the CBM and VBM lie at the $\Gamma$ -point. In heavily doped or oxygen deficit semiconductors the chemical potential is shifted, so that bands cross the Fermi level. In this case we determined the effective band masses $m_{B,i}$ from bands states close to the Fermi level. In order to verify the large effective mass in experiments, we treat both defects separately and use the 300meV Oxygen deficit band level ($SrTiO_{3-v}$ v=0.98) as donor state and the high effective mass $m^*/m_0 > 12$ of the Nb-doped case $SrTi_{1-x}Nb_xO_3$ x=0.02. This separation can explain the interaction of both defects with smaller concentration than calculations can perform.

**Table 3** Calculated effective mass of several $SrTiO_3$-related perovskite phases

| Compo­sition | Lattice constants $a_o$ [nm] | Magnetic moment | Effective band mass $m_{B,I}/m_0$ at $\Gamma-\Delta$ | Effective mass $m^*/m_0$ |
|---|---|---|---|---|
| *A*-site substitution | | | | |
| $CaTiO_3$ | 0.3795 | 0 | 4.0 | 4.0 |
| $SrTiO_3$ | 0.3905 | 0 | 4.4 | 4.8 |
| $BaTiO_3$ | 0.4014 | 0 | 4.8 | 5.3 |
| *B*-site substitution | | | | |
| $SrVO_3$ | 0.3842 | 0.13 | 8.0 | 10.0 |
| $SrTiO_3$ | 0.3905 | 0 | 4.4 | 4.8 |
| $SrMoO_3$ | 0.3965 | 0.01 | 4.0 | 3.2 |
| $SrNbO_3$ | 0.4112 | 0 | 3.0 | 2.7 |
| $SrZrO_3$ | 0.4093 | 0 | 2.0 | 1.7 |
| $SrHfO_3$ | 0.4114 | 0 | 1.3 | 1.1 |
| $SrMnO_3$ | 0.3821 | 3.0 | 0.3 | 0.2 |
| $SrFeO_3$ | 0.3851 | 3.27 | 0.3 | 0.1 |
| $SrCoO_3$ | 0.3835 | 2.28 | 0.3 | 0.2 |
| $SrCuO_3$ | 0.3810 | 0 | 0.3 | 0.1 |
| *A*-and *B*-site substitution | | | | |
| $NaTaO_3$ | 0.3931 | 0 | 8 | 1.4 |
| $KTaO_3$ | 0.3988 | 0 | 6 | 7.2 |
| $Ca_{0.5}TaO_3$ | 0.3875 | 0 | 0.18 | 0.06 |

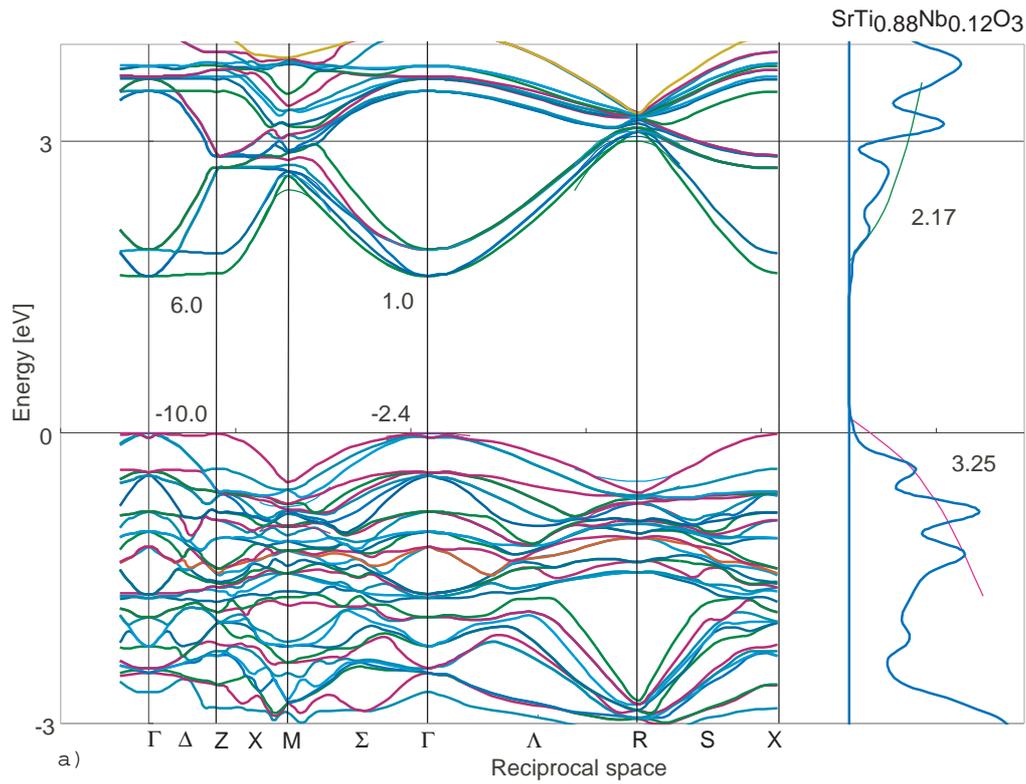

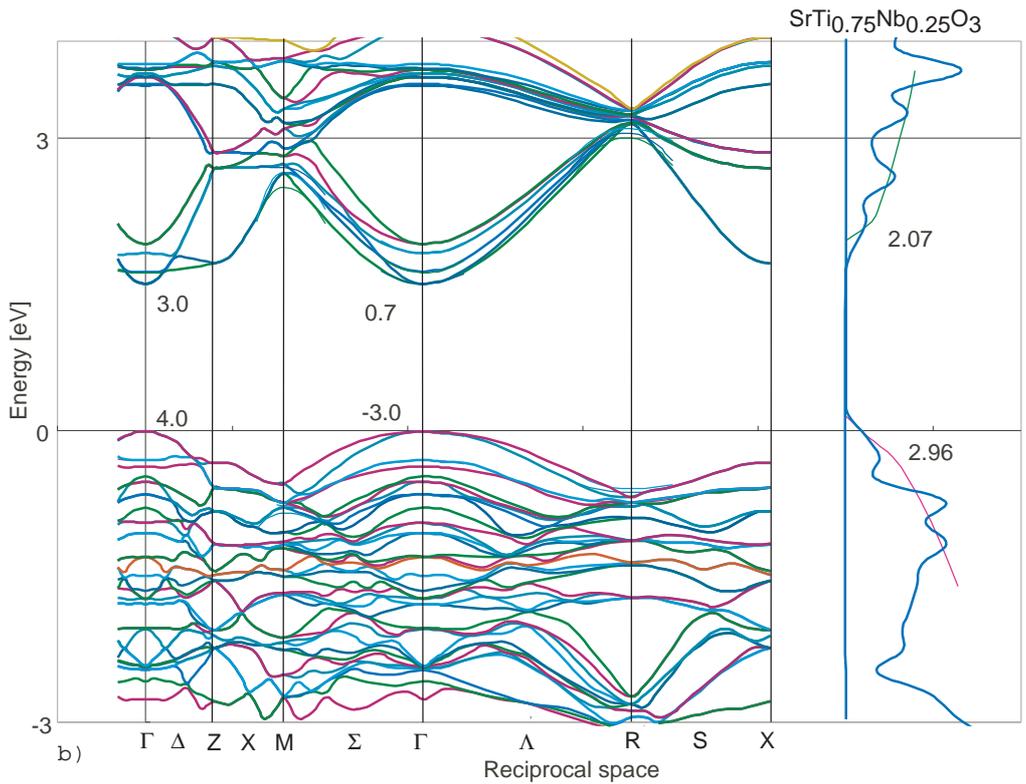

**FIG 8.** Electronic Bandstructure of (a) $SrTi_{0.88}Nb_{0.12}O_3$, (b) $SrTi_{0.75}Nb_{0.25}O_3$.

Summarizing all results, four effects can be distinguished for doping with heavier elements in $SrTiO_3$, namely the enlargement of the atomic mass, the shift of the chemical potential, the enlargement of the lattice constants according to Vegards' rule and Oxygen reduction due to the increase in the number of electrons. The enhanced calculated mass can be explained by the higher atomic mass in the order Ba, Sr and Ca, while for the *B*- site doping the shift in the chemical potential causes an opposite behavior. Additionally, in the case of most transition metals other than Ti on the *B*- site like V, Mn, Fe and Co, spin splitting occurs, which leads to a magnetic moment. The reduced symmetry due to doping atoms causes the splitting of degenerate bands which increases the number of calculated bands.

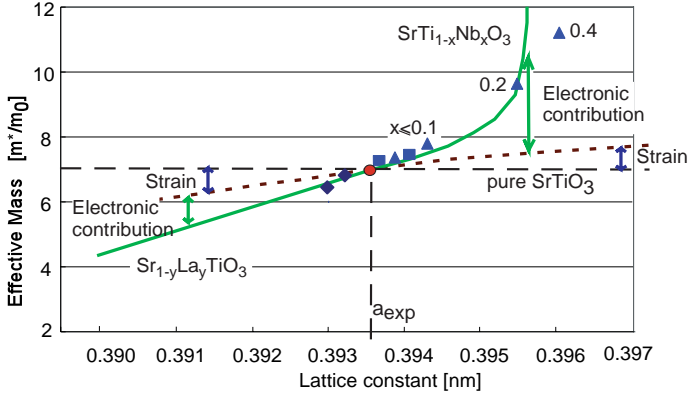

**FIG. 9.** Effective electron mass $m^*$ as a function of isotropically expanded lattice constants for Nb- and La-doped SrTiO$_3$. (squares: experimental data on single crystals from ref. 1, triangles: data on thin films from ref. 2.) The bright line refers to calculation results from Fig.. 5(c) and Fig. 7 and the dotted line from Fig.. 3(d).

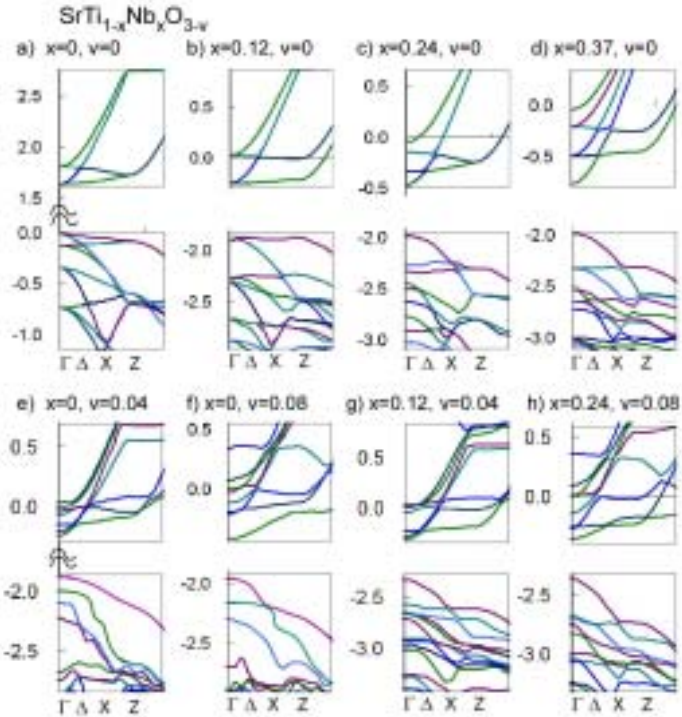

**FIG. 10.** Electronic bandstructure near the band gap at the Γ-point for SrTi$_{1-x}$Nb$_x$O$_{3-v}$ with different concentration of defects, a) pure SrTiO$_3$ for comparison, b-d) Ti substitution by Nb, e-f) Oxygen vacancy, g-h) Ti-substitution by Nb with O-vacancy.

## Discussion

The simulations on the SrTiO$_3$ and related single crystals showed a good agreement to experiments and show the validity of the effective mass concept, which is a simplified parameterization of the complex interaction between traveling electrons in the lattice. With Nb-doping the effective mass increases, whereas with La-doping it decreases. If we consider this as a general rule, it means that electron donation on the B- site leads to higher effective mass, while on the A-

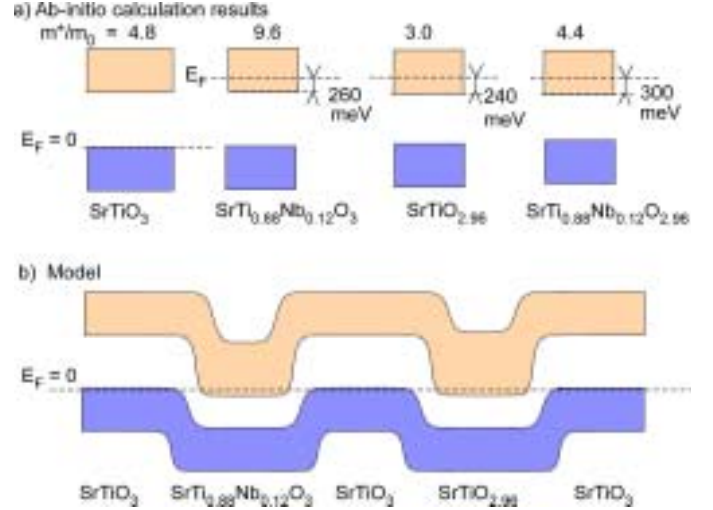

**FIG 11.** According to the a) calculation results for defects, the b) model of the electronic bandstructure of SrTi$_{1-x}$Nb$_x$O$_{3-v}$ with diluted doping atoms and O-vacancies was deduced.

site it leads to smaller effective mass. Considering the substitution, the reverse behavior of A- and B-site is also obvious, as for substitutions of elements in the same row of the periodic systems A=Ca, Sr, Ba $m^*/m_0$ increases; these were calculation result shown in table 3, and meanwhile experimentally verified10. On the other hand, for B=Ti, Zr, Hf it decreases (table 3). In other words, in the case of electron injection by doping or a smaller orbital radius with heavier elements, both mechanisms lead obviously to the same result, with opposite effect for the A- or B- site. The increase in the effective mass in the case of BaTiO$_3$ leads to an increase in the Seebeck coefficient, which has indeed been found experimentally for Sr$_{0.9-z}$Ba$_z$La$_{0.1}$TiO$_3$ solid solutions.10 Although negligible at the considered temperatures, an increase in $m^*$ also leads to a reduction of the electronic part of the thermal conductivity, resulting in the maximum achievable dimensionless figure of merit $ZT$=0.3.10,42

We showed, that the preliminary discrepancy between the experimental effective mass of $m^*/m_0$ =7.21,2 and the calculated one of $m^*/m_0$ = 4.8 almost disappears, because of several factors, namely, lattice expansion, oxygen deficit, doping and their interaction increases the effective mass. Other explanations for the increase in the effective mass caused by interactions with other electrons, atoms or the lattice, such as (a) relativistic effects in the vicinity of heavy atom cores, which might be responsible for the hump in the band-structure (Fig. 1), (b) an additional self-energy, (c) electron correlation effects, (d) charge density waves or (e) electron phonon coupling via polarons are some of the effects most commonly discussed in the literature.34 The investigations in this study showed that perovskite materials still have a potential for increased effective mass, which would further improve the Seebeck coefficient and other thermoelectric properties. This study focused on lattice strain and doping, and in a subsequent paper the effective mass of layered perovskites is reported7.

## Summary

The thermoelectric power-factor can be improved, if the effective mass is increased, because the gain in Seebeck coefficient $S^2$ is larger than the decrease in the mobility. Using ab-initio calculations the effective mass of SrTiO$_3$ was calculated as $m^*/m_0$=4.8 which agrees well with experimental values. Lattice expansion increases the effective mass, but this increase is usually much less than that caused by electronic contributions due to doping or vacancies. Whether the doping of electron donors occurs on the *A*- or *B*-site, it leads to opposite behavior concerning the effective mass, namely an increase for x for SrTi$_{1-x}$Nb$_x$O$_3$ and a decrease for y for Sr$_{1-y}$La$_y$TiO$_3$. The increase for Nb-doped SrTiO$_3$ can be explained by the flattening of the *e2g*- band. With increasing doping concentration at x>0.25 the heavy and light bands at the Γ point spilt in energy, which leads to a dominance of the light electrons. For Sr$_{1-y}$La$_y$TiO$_3$ this splitting also occurs at y>0.25. Finally, the general finding of this study is, that the effective mass $m_{DOS}$ derived from the density of states corresponds well to the average over effective masses of different electronic bands $m_{Bis}$ but the effective mass $m^*$ relevant for calculating the electrical conductivity requires detailed deduction from the bands close to the bandgap and the separation between doping and oxygen vacancies.

**Acknowledgement**

Useful discussions with Shingo Ohta of Nagoya University are gratefully acknowledged.